\documentclass[12pt]{article} 

\usepackage{latexsym} 
\usepackage{amssymb}  
\usepackage{epsfig}       

\newcommand{\beq}{\begin{equation}}
\newcommand{\eeq}{\end{equation}}
\newcommand{\ba}{\begin{array}}
\newcommand{\bea}{\begin{eqnarray}}
\newcommand{\ea}{\end{array}}
\newcommand{\eea}{\end{eqnarray}}

\newcommand\comment[1]{ \hbox{[{\it Comment suppressed here.}\/]} }
\newcommand\hide[1]{}

\newcommand{\skipover}[1]{}

\makeatletter 

\def\appendix{\par                              
    \setcounter{section}{0}                     
    \setcounter{subsection}{0}
    \renewcommand{\theequation}{\Alph{section}.\arabic{equation}}
    \renewcommand{\thesection}{Appendix \Alph{section}
                \setcounter{equation}{0}  } 
}

\def\applabel#1{\@bsphack
  \protected@write\@auxout{}%
         {\string\newlabel{#1}{{\Alph{section}}{\thepage}}}%
  \@esphack}

\def\section{
\setcounter{equation}{0}        
\@startsection {section}{1}{\z@}{-3.5ex plus -1ex minus 
 -.2ex}{2.3ex plus .2ex}{\large\bf}}
\renewcommand{\theequation}{\arabic{section}.\arabic{equation}}

\def\subsection{\@startsection{subsection}{2}{\z@}{-3.25ex plus -1ex minus 
 -.2ex}{1.5ex plus .2ex}{\normalsize\bf}}

\def\subsubsection{\@startsection{subsubsection}{3}{\z@}{-3.25ex plus
 -1ex minus -.2ex}{1.5ex plus .2ex}{\normalsize}}

\makeatother   

\newsavebox{\eqlabel}

\makeatletter  
\newlength{\numblen}
\newsavebox{\eqnumb}
\def\@eqnnum{\savebox{\eqnumb}{\rm (\theequation)}%
\settowidth{\numblen}{\usebox{\eqnumb}}%
\makebox[\numblen][l]{\usebox{\eqnumb}~~~\usebox{\eqlabel}}}
\makeatother   

\newenvironment{equationwithlabel}[1]{ %
  \savebox{\eqlabel}{#1}
  \begin{equation}\label{#1} }{\end{equation}} 
\newcommand{\beql}[1]{\begin{equationwithlabel}{#1}}
\newcommand{\eeql}{\end{equationwithlabel}}

\begin{document}

\title{\vspace{-0.5in}\bf Collision--Induced 
Decay of Metastable Baby Skyrmions}

\author{Daniel A. Dwyer\\
{\normalsize Department of Physics, Massachusetts Institute of Technology}\\
{\normalsize Cambridge, MA 02139 }\\[1.0ex]
Krishna Rajagopal \\
{\normalsize Center for Theoretical Physics, 
Massachusetts Institute of Technology}\\
{\normalsize Cambridge, MA 02139 }
}

\newcommand{\preprintno}{
  \normalsize MIT-CTP-2970
}

\date{April 13, 2000, \preprintno}

\begin{titlepage}
\maketitle
\def\thepage{}          

\begin{abstract}
Many extensions of the standard model predict heavy metastable
particles which may be
modeled as solitons (skyrmions of the Higgs field), 
relating their particle number to a winding number. 
Previous work has shown that
the electroweak interactions admit processes in which
these solitons decay, violating standard model baryon number.  
We motivate the hypothesis that baryon--number--violating
decay is a {\it generic} outcome of collisions between these heavy
particles. We do so by exploring a 2+1 dimensional theory 
which also possesses metastable skyrmions.
We use relaxation techniques to determine the size, shape and energy of 
static solitons in their ground state.  
These
solitons could decay by quantum mechanical
tunneling. Classically, they are metastable: only
a finite excitation energy is required to induce their decay.  
We attempt to induce soliton decay in a classical simulation
by colliding pairs of solitons.  We analyze the collision
of solitons with varying
inherent stabilities and varying incident velocities and
orientations.
Our results suggest that winding-number violating decay
is a generic outcome of collisions. All that 
is required is sufficient (not necessarily very
large) incident velocity; no fine-tuning of initial conditions is required.
\end{abstract}

\end{titlepage}

\renewcommand{\thepage}{\arabic{page}}

\section{Introduction}

\subsection{Motivation}

Many extensions to the standard model which involve  
strong dynamics at the electroweak
scale include new heavy particles
which have been modeled as solitons.  The
simplest model within which such 
particles can be analyzed is the standard electroweak theory
with the Higgs boson mass $m_H$ taken to infinity and
with a Skyrme term \cite{Skyrme} added to the Higgs sector.
With these modifications, the Higgs sector supports
a classically stable soliton whose mass is of order
the weak scale, typically a few TeV.\cite{Gipson}  

To understand how solitons arise, note that in the absence of the weak gauge
interactions, the Higgs sector of the standard model
is a four-component scalar field theory in which
a global $O(4)$ symmetry is spontaneously broken to $O(3)$, with
vacuum manifold $S^3$.
In the $m_H\rightarrow \infty$ limit, the dynamics
is that of an $O(4)$ nonlinear sigma model.  
Field configurations are maps from three
dimensional space onto $S^3$, and the solitons
(skyrmions) are configurations which
carry the associated winding number.  The winding number 
is topological and soliton number is 
conserved.  

Gauging the weak interactions changes the
picture qualitatively 
because the winding number of the 
Higgs field is not invariant under large gauge transformations.
This means that a soliton can either be described as a
skyrmion of the Higgs field with gauge field $A_\mu=0$ or, equivalently, 
as a topologically trivial Higgs field configuration with a suitably
chosen nonvanishing $A_\mu$.
The latter description makes manifest the fact that 
there are sequences of gauge and Higgs field
configurations, beginning with a soliton and ending with
a vacuum configuration, such that all configurations in
the sequence have finite energy.  This means that
the soliton is only metastable: it is separated from
the vacuum only by a finite energy barrier and can
decay quantum mechanically by 
tunneling.\cite{DhokerFarhi,AmbjornRubakov,RST,EilamStern}
Or, the soliton can be kicked over the barrier
if it is supplied with energy. The process in which an 
electroweak soliton 
is hit with a classical gauge field pulse (a coherent state
of $W$-bosons) and caused to decay has been analyzed 
numerically.\cite{FGLR}
It is even possible to find a limiting case of the theory in 
which the
quantum mechanical cross-section
for a process in which a soliton is struck by a single
$W$-boson and induced to decay can be calculated analytically.\cite{FGLR}
In any process in which a soliton is destroyed, one net baryon
and one net lepton from each standard model generation is anomalously
produced.\cite{FGLR}

Electroweak solitons have also been studied in the electroweak
theory with finite Higgs mass, in which the Higgs
sector is a linear sigma model.\cite{Baacke}   If a Skyrme
term is added to the theory, metastable electroweak solitons
exist if $m_H$ is sufficiently large.
In the linear sigma model, the Higgs field can
vanish at a point in space with only finite cost in energy.
The Higgs winding number is therefore not topological even
in the absence of gauge interactions.  This means that in
a world with gauge interactions and a finite Higgs mass,
there are two ways for solitons to decay: either via nontrivial
gauge field dynamics, as sketched in the previous paragraph, or via the Higgs
field itself simply unwinding.\cite{LueTrodden}

The metastable electroweak soliton 
is an intriguing object to study. And yet, it is not found
in the standard electroweak theory where the Higgs sector is
a linear sigma model with no higher derivative terms.   
The Higgs sector of the standard model
is best thought of as an effective field theory
describing the low energy (weak scale) dynamics of the light
degrees of freedom in 
some higher energy theory. The simplest examples of higher
energy theories which feature particles which can be described
as electroweak solitons in the low energy theory are technicolor theories,
in which the technibaryons play this role.
Regardless of whether the underlying theory is
specifically a technicolor model, it will introduce all 
higher derivative terms allowed by symmetries, 
including the Skyrme term, into the Lagrangian of the
low energy effective theory.  If the Higgs boson is discovered to be 
light (say, with mass $m_H\lesssim v=250$ GeV), 
the correct low energy
effective field theory will almost certainly not support solitons,
regardless of the physics of the higher derivative terms.
If the Higgs boson is discovered to be heavy,
there will be some class of appropriate high energy theories 
whose low energy effective field theories,
although more complicated than that obtained simply by adding
a Skyrme term to the standard model, feature metastable 
electroweak solitons.  
Discovery of the corresponding TeV scale particles would confirm
that nature chooses such a theory.

Processes in which two metastable electroweak solitons collide have 
to date not been
studied.  Our purpose in this paper
is to use the analysis of a two dimensional
toy model which shares some (but not all) of the features
outlined above to motivate the hypothesis that
the generic outcome 
of such collisions may be the destruction of one or both
solitons. This suggests (but certainly does not demonstrate) that
baryon number violation is the generic outcome of 
collisions between two of the TeV scale particles which
can be modeled as solitons.

As a sideline, we note that our numerical methods work equally
well for describing soliton--soliton and soliton--antisoliton
collisions.  Our focus is on soliton decay in soliton--soliton
collisions; we note, however, that 
the numerical simulation of soliton--antisoliton annihilation
in the Skyrme model
is well-known as a difficult numerical problem, plagued
with instabilities.\footnote{See Ref. \cite{Koonin} for
classical simulations of skyrmion--skyrmion scattering
in the 3+1 dimensional Skyrme model
which report instabilities in the simulation of 
skyrmion--antiskyrmion annihilation; see Ref. \cite{Crutchfield} 
for a discussion of the origin of the instabilities and 
Refs.\cite{Crutchfield,Halasz} for efforts
to overcome them.}  
We are able to follow
soliton--antisoliton annihilation without difficulty (with
energy conserved at the part in $10^4$ level). This suggests
that our numerical methods --- in particular 
the use of the linear sigma model ---
may be of broad utility when generalized to 3+1 dimensions. 

\subsection{Metastable Baby Skyrmions}

Let us now introduce the 2+1 dimensional model whose
metastable solitons we analyze. The Lagrangian density, which
describes the dynamics of a 
three
component scalar field $\vec\phi =  (\phi^1,\phi^2,\phi^3)$, is
\begin{eqnarray}
{\cal L} = F\Biggl[\frac{1}{2}\partial_\alpha\vec{\phi}\cdot
\partial^\alpha\vec{\phi} &-& \frac{\kappa^2}{4}
(\partial_\alpha\vec{\phi}\times\partial_\beta\vec{\phi})\cdot
(\partial^\alpha\vec{\phi}\times\partial^\beta\vec{\phi})\nonumber\\ 
&-& \mu^2(v-\vec{n}\cdot\vec{\phi}) - 
\lambda\left(\vec{\phi}\cdot\vec{\phi} - v^2\right)^2\Biggr]\ .
\label{lagrangian}
\end{eqnarray}
Here, $\vec n$ is a unit vector which we choose to be $(0,0,1)$.

To understand the features of this Lagrangian, it is worth
beginning by setting $\mu^2=0$ and taking the limit 
$\lambda\rightarrow \infty$.
When $\mu^2=0$, the theory has an $O(3)$ symmetry. 
For $\lambda\rightarrow \infty$, one removes the fourth term from 
(\ref{lagrangian}) and instead imposes the constraint that 
$\vec{\phi}\cdot\vec{\phi} = v^2$ at all points in space and time.
Because the field $\vec{\phi}$ is constrained to take values on a two-sphere
of radius $v$, field configurations with fixed boundary conditions at
infinity can be classified by their winding number 
\begin{equation}
Q = \frac{1}{8\pi v^3}\int{\epsilon_{ab}\vec{\phi}\cdot
(\partial_a\vec{\phi}\times\partial_b\vec{\phi})d^2x} = 
\frac{1}{4\pi v^3}\int{ \vec{\phi}\cdot
(\partial_x\vec{\phi}\times\partial_y\vec{\phi})\,dx\,dy} \ ,
\end{equation}
which is integer-valued and topological:  configurations with 
different winding
number cannot be continuously deformed into one 
another.   
This suggests the possibility of soliton solutions to
the classical equations of motion.  Solitons in 2+1--dimensional
$O(3)$ sigma models were first discussed in Ref. \cite{BelavinPolyakov},
and their quantum field theoretic properties were analyzed in
Refs. \cite{WilczekZee,MackenzieWilczek}.  Such solitons
are often called baby skyrmions \cite{MackenzieWilczek} 
because of their similarity to 3+1--dimensional skyrmions.
Although our
motivation is the analogy to 3+1--dimensional electroweak
solitons, we note that baby skyrmions themselves do
arise in certain 2+1--dimensional electron systems which
exhibit the quantum hall effect \cite{Sondhi}, although
the Lagrangian used in their description differs from 
that in Eq. (\ref{lagrangian}).

The four-derivative term in the Lagrangian (\ref{lagrangian}) 
is the analogue of the Skyrme term.
It stabilizes putative solitons against shrinking
to arbitrarily small size.  If we were working in three spatial dimensions,
the two-derivative term would stabilize putative solitons against
growing to arbitrarily large size.
In two spatial dimensions, however, the two-derivative term
cannot play this role because its
contribution to the energy of a configuration 
is scale invariant.
We must therefore introduce a zero-derivative term 
in order to stabilize
solitons against growing without bound.  Such a term must explicitly
break the $O(3)$ symmetry, and therefore has no
analogue in 3+1--dimensional electroweak physics, in which
no explicit $O(4)$ symmetry breaking terms are allowed.
The particular form of the $\mu^2$ term in (\ref{lagrangian}) 
therefore has no 
electroweak motivation; it 
is analogous to a pion mass term
in the 3+1 dimensional Skyrme model, but this is not relevant to us.   
This model (with $\mu^2$ nonzero and $\lambda\rightarrow\infty$)
was considered in Ref. \cite{Piette1}, and its solitons
have been analyzed in
detail in Refs. \cite{Piette2,Piette3}. 
Similar models, differing only in the choice of the explicit
symmetry breaking term in the Lagrangian, have also been 
analyzed.\cite{OtherPotentials}  

The soliton mass
and size in the theory with Lagrangian (\ref{lagrangian})
with $\lambda=\infty$ are given by\cite{Piette3}
\begin{equation}
M_{\rm sol} = 19.47 F \left[ a_1 \sqrt{\frac{\kappa\mu}{0.316}} 
+ a_2 \right]\ ,
\ \ \ \ \ \ R_{\rm sol}\sim (3-4) \kappa \sqrt{\frac{0.316}{\kappa\mu}} \ ,
\label{massradius}
\end{equation}
with $a_1$ and $a_2$ dimensionless constants (independent
of $\kappa\mu$)  satisfying $a_1+a_2=1$.
The parametric dependence of these results can be understood
by 
noting that the energy of a configuration of size $R$ 
receives contributions of order $FR^0$, $F\kappa^2 R^{-2}$ and $F\mu^2R^2$
from the first three terms in the Lagrangian (\ref{lagrangian})
and that, as described
above, a soliton is stabilized by the balance between
the four-derivative $\kappa^2$ term  and the zero-derivative $\mu^2$ term.

If we stopped here, with $\lambda$ infinite, our solitons would
be absolutely stable, rather than metastable.  
Soliton--soliton
collisions have been simulated in this theory, but of course
the solitons never decay.\cite{Piette3}  
Once $\lambda$ is finite, the fields are allowed
to deviate from $\vec{\phi}\cdot\vec{\phi} = v^2$, and the soliton
configuration with $\vec{\phi}\cdot\vec{\phi} = v^2$ found previously
in the $\lambda\rightarrow\infty$ theory may unwind and decay.  Indeed, 
we will see that soliton solutions do not exist for $\lambda$
less than some $\lambda_c$.  If $\lambda>\lambda_c$, metastable
solitons exist: these solitons are classically stable if left unperturbed, but
can be induced to decay if supplied
with sufficient energy. Our goal is to determine whether the
means by which the energy is delivered is important  
or whether soliton decay is the result of generic soliton--soliton
collisions, without finely tuned 
initial conditions.

For our purposes, $\lambda$ is the most
important parameter in the theory because by choosing its value,
we control the energy required to make the soliton
decay and indeed control whether solitons exist in the
first place.  We are not interested in the dependence
on the other parameters, and indeed most of them can
be scaled away.  We first set $v=1$ by rescaling $\phi$.
Next, the constant $F$ has units of energy and we henceforth
measure energy in units such that $F=1$.  
Next, $\kappa$ has units of length
and we henceforth measure length in units such that $\kappa=1$.
Note that this means that $\hbar\neq 1$ in our units, but
this will not concern us as we only discuss the classical
physics of this model.  We have set the speed of light $c=1$ throughout.
The parameters $\mu^{-1}$ and $\lambda^{-\frac{1}{2}}$ are also
length scales in
the Lagrangian, and the theory is 
therefore fully
specified by the two dimensionless parameters $\lambda\kappa^2=\lambda$
and $\mu\kappa=\mu$.  Although results do depend on $\mu$,
we are not interested in this dependence, and we choose
to follow Ref. \cite{Piette3} and set $\mu^2=0.1$ throughout.
Once we have chosen units with $F=\kappa=1$ and have chosen
to set $\mu^2=0.1$, then $M_{\rm sol}=19.47$ and $R_{\rm sol}\sim (3-4)$
in the theory with $\lambda=\infty$.

In Section 2, we find metastable soliton configurations
for finite values of $\lambda$ with $\lambda>\lambda_c\sim 7.6$.
We shall see that for all values of $\lambda$ for which solitons
exist, the soliton mass and size change little from their values
at $\lambda\rightarrow\infty$.  Although we do
not explore their dependence on $\mu$, we expect
it would be similar to that in (\ref{massradius}).
In Section 3, we present our results on soliton--soliton collisions.
We find that soliton decay occurs for incident
velocities greater than some critical value $v_c$. 
We explore how this critical velocity depends on $\lambda$
and on the initial impact parameter and relative orientation of
the two solitons. We find that $v_c$ is less than or of order half
the speed of light regardless of the relative orientation
as long as $\lambda \lesssim 2 \lambda_c$ and 
$b$ is less than or of order the soliton size.  
Thus, inducing soliton decay does not require specially
chosen initial conditions; it is a generic outcome of soliton--soliton
collisions. 
We make concluding remarks in Section 4.\cite{thesis}

It perhaps goes without saying that our model is at best a crude
toy model for the electroweak physics which motivates our
analysis.  First, we work in 2+1 dimensions.
Second, in order for the theory to have soliton
solutions we are forced to include a
zero-derivative explicit
symmetry breaking term not present in the electroweak theory.
Third, we do
not introduce a gauge field. Hence, our solitons can
only decay via unwinding the scalar field;  in the electroweak
theory, gauge field dynamics introduces a second decay mechanism
which has no analogue in our theory.
Related to this, our solitons
are absolutely stable for $\lambda=\infty$, whereas electroweak
solitons are metastable even for $m_H=\infty$. This is 
perhaps the biggest qualitative difference between our
model and electroweak physics.  Fourth, one may 
worry that even if an analysis along the lines of ours
were done in the 3+1 dimensional electroweak theory itself, the
momenta required would make it impossible to analyze soliton
decay within the effective theory.
This concern may be evaded
for solitons which are almost unstable: in this circumstance,
for example, $W$-soliton collisions can result in soliton
destruction even if the $W$-boson momentum is small enough
that the calculation is controlled.\cite{FGLR}  
Soliton--soliton scattering in
our model is far from being a complete analogue
of the scattering of TeV scale particles which
can be modeled as metastable electroweak solitons;  
we nevertheless hope that our central result, namely
that metastable baby skyrmions in 2+1 dimensions are destroyed
in collisions with generic initial conditions, motivates
future work on baryon number violating scattering in
this sector.

\section{Finding Static Solitons}

Before we can study soliton--soliton collisions, we must
find the metastable soliton configurations for different
values of $\lambda$.  We do this by looking for configurations
which 
minimize the static
Hamiltonian $H_{\rm static}$ at a given $\lambda$. 
The static Hamiltonian is given by
\begin{equation}
H_{\rm static} = -\int d^2x \, {\cal L}_{\rm static}\ ,
\end{equation}
where ${\cal L}_{\rm static}$ is the Lagrangian density of 
(\ref{lagrangian}) with all terms containing time derivatives
set to zero.

We discretize $H_{\rm static}$ on a square lattice of $125\times 125$
points, with the spatial separation between points given by 
$\Delta x=0.2$ (in our units in which $\kappa=1$).  
We discretize the two derivative term
in the standard fashion, writing it as a sum over terms
like 
\begin{equation}
\left[\frac{\phi^i(x,y)-\phi^i(x-\Delta x,y)}{\Delta x}\right]^2\ .
\end{equation}  
The Skyrme term
is trickier to handle, because it involves terms like
\begin{equation}
\partial_x \phi^1 \,\partial_y \phi^1 \,\partial_x \phi^2 
\,\partial_y \phi^2\ .
\end{equation}
We discretize this contribution to the Hamiltonian as a sum over terms like
\begin{eqnarray}
&\ &\left(\frac{  \phi^1(x+\Delta x,y)-\phi^1(x-\Delta x,y) }{2\Delta x}\right)
\left(\frac{  \phi^1(x,y+\Delta x)-\phi^1(x,y-\Delta x) }{2\Delta x}\right)
\nonumber\\
&\ &\times
\left(\frac{  \phi^2(x+\Delta x,y)-\phi^2(x-\Delta x,y) }{2\Delta x}\right)
\left(\frac{  \phi^2(x,y+\Delta x)-\phi^2(x,y-\Delta x) }{2\Delta x}\right)
\qquad\quad
\end{eqnarray}
In this way, we ensure that within each term in the sum over lattice sites,
all spatial derivatives are centered at the same point in space.
Discretizing the Hamiltonian in this fashion ensures that
discretization errors are of order $(\Delta x)^2$.

In order to find a soliton, we begin with a guess (which
we describe momentarily) for 
the configuration $\vec \phi(x,y)$ and perform a numerical minimization
of the static Hamiltonian using the conjugate gradient
method of Ref. \cite{Press}. (It is important
to use a method such as this one, which minimizes
a function of $N$ variables using computer memory of order $N$ rather
than of order $N^2$ since we have an $N=3\times 125 \times 125$ dimensional
configuration space.)
In order to minimize the
energy, the conjugate gradient routine needs
expressions for the gradient
of the energy at any point in our $N$ 
dimensional configuration space, with respect to
each direction in this configuration space.
We obtain these expressions
by varying the discretized $H_{\rm static}$ 
with
respect to the $\phi^i$ at each lattice site. 
(These expressions will of course also 
appear as the terms with no time derivatives in the dynamical
equations of motion of Section 3.)

For $\lambda\rightarrow\infty$, soliton solutions can
be written in the form\cite{Piette1,Piette2}
\begin{equation}
\vec{\phi}(r,\theta) =\left( \begin{array}{c}
                                        \sin f(r) \cos \theta\\ 
                                        \sin f(r) \sin \theta\\ 
                                        \cos f(r)
                                \end{array} \right)
\label{guess}
\end{equation}
where $f(r)$ satisfies the following conditions:
\begin{eqnarray}
f(0)=\pi, \\
\lim_{r\rightarrow\infty}f(r)=0\ .
\end{eqnarray}
(We define polar coordinates such that the soliton
is centered at $r=0$, $\theta=0$ is the positive $y$-axis,
and $\theta$ increases in a clockwise direction.)
Note that because of the $\mu^2$ 
term in the Lagrangian which breaks the $O(3)$ symmetry,
$\vec \phi$ must point in the $\phi^3$ direction at large $r$.
The $O(2)$ symmetry associated with rotations in the $(\phi^1,\phi^2)$
plane is not broken in the Lagrangian; in the solution,
these rotations are mapped onto rotation in the $(x,y)$
plane about the soliton center.
This configuration
is thus a two-dimensional analogue of what in three dimensions
is called a hedgehog configuration.

\begin{figure}[thb]
\begin{center}
\begin{tabular}{cc}
\epsfig{file=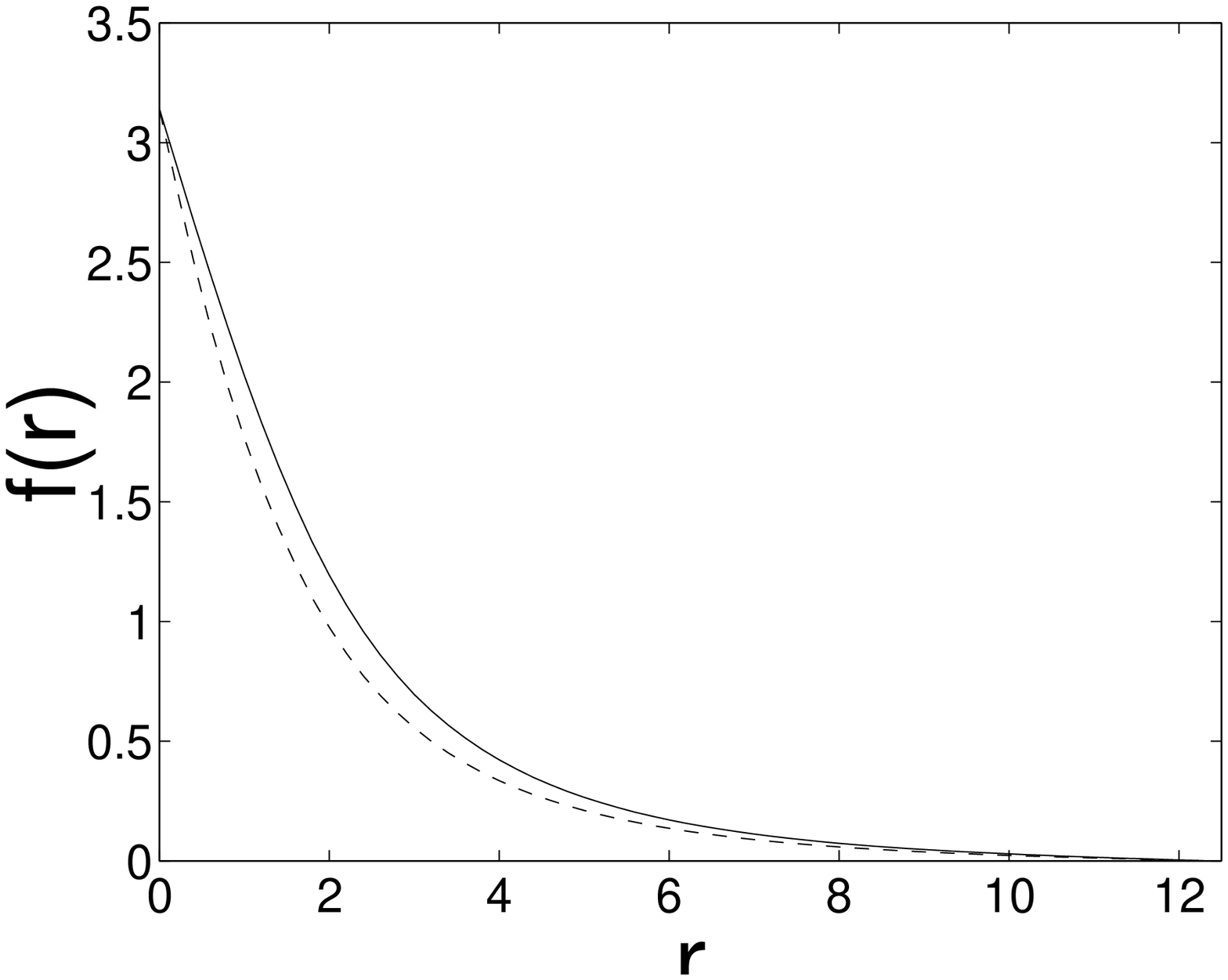,width=2.7in} &
\epsfig{file=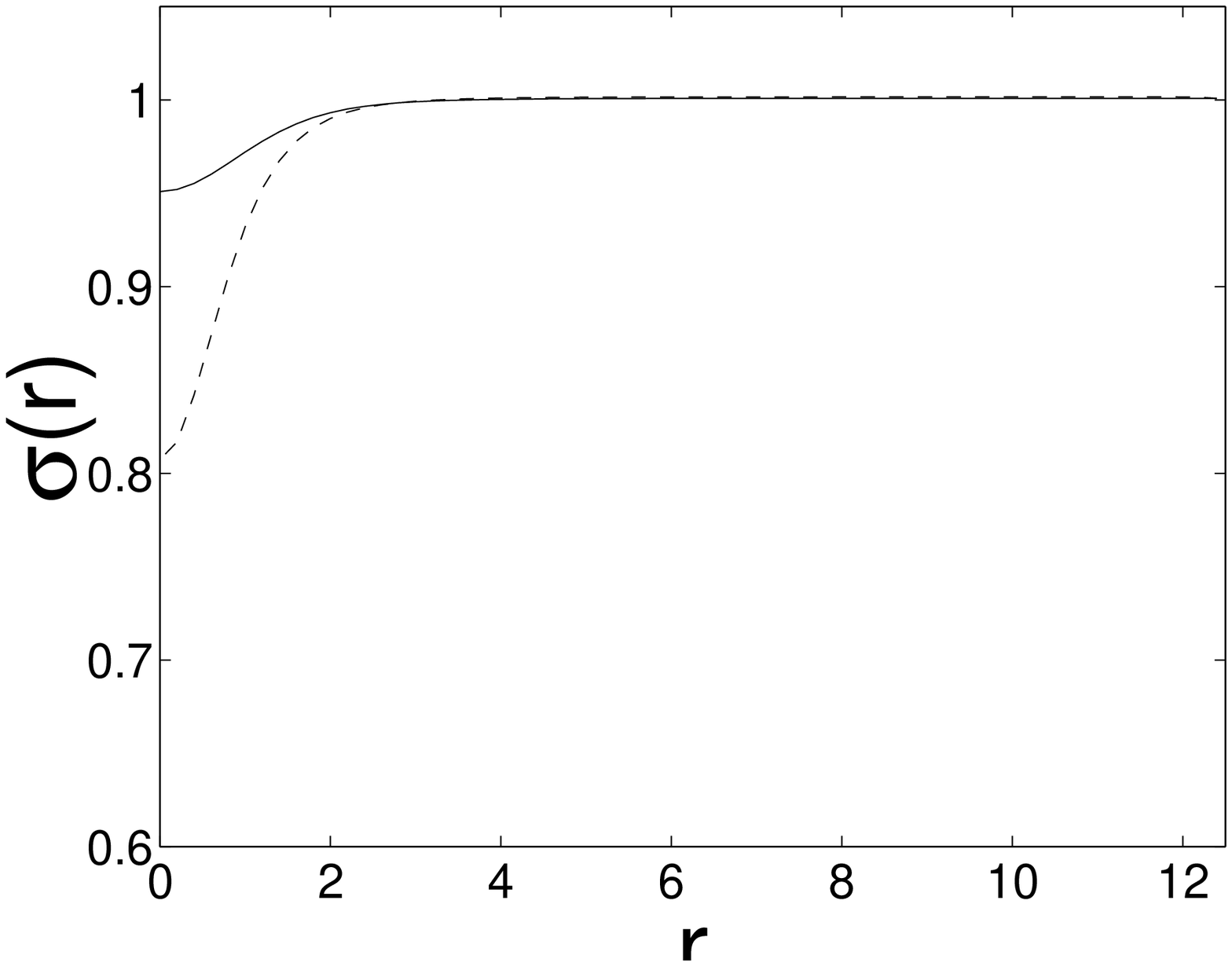,width=2.7in} \\
\epsfig{file=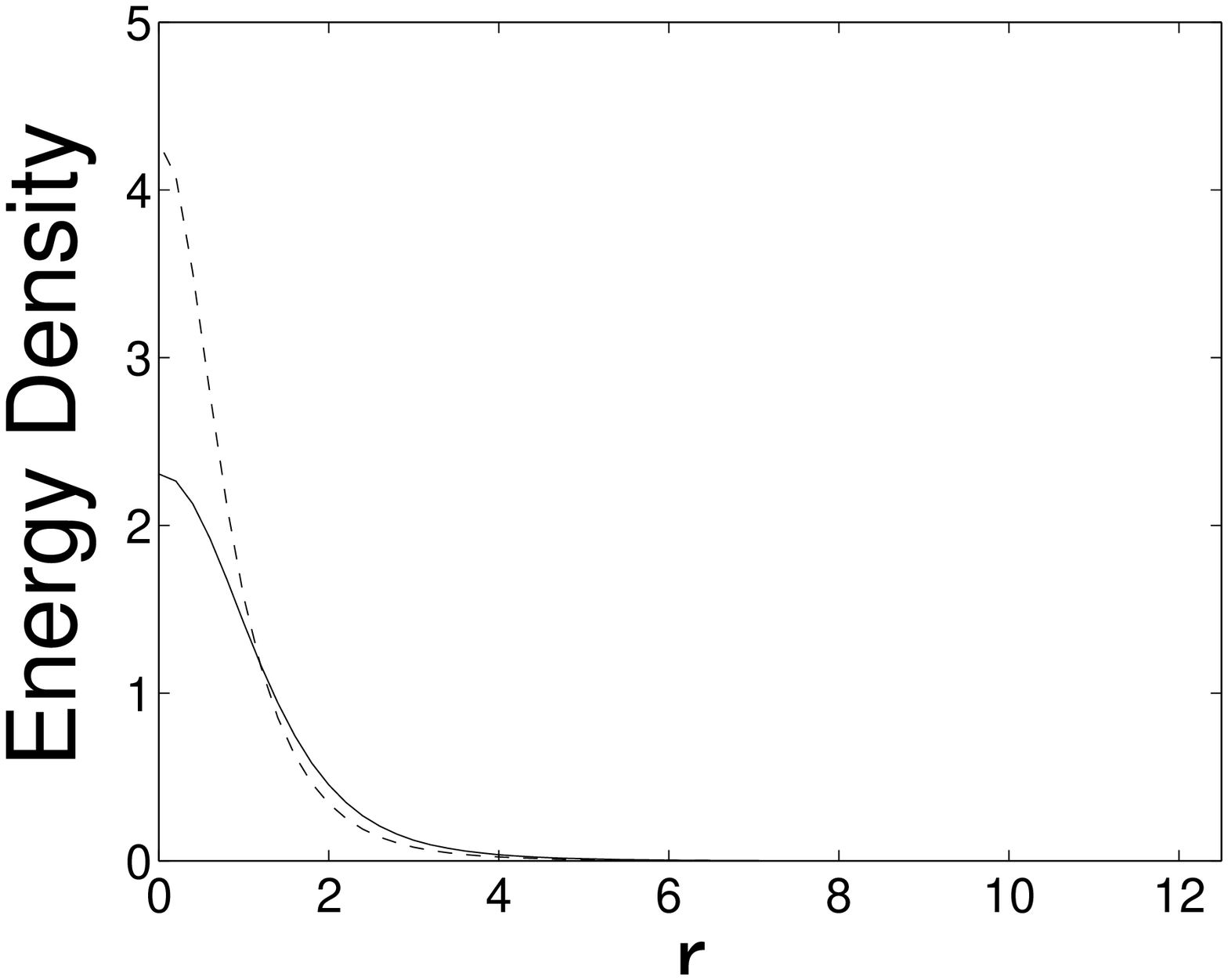,width=2.7in}
\end{tabular}
\caption{$f(r)$, $\sigma(r)$ and the energy density for the
solitons with $\lambda=15$ (solid curves) and $\lambda=7.7$ (dashed
curves).}
\label{fig:solitonprofiles}
\end{center}
\end{figure}

In our search
for solitons at finite $\lambda$, we therefore begin by choosing 
a reasonably large $\lambda$, namely $\lambda=15$, and making
an initial guess of the form (\ref{guess}) with $f(r)=\pi\exp(-r/2)$.
We then run the conjugate gradient relaxation algorithm
repeatedly, until the change in the energy between 
successive relaxation steps is smaller than one part 
in $10^{10}$.\footnote{As a check, 
we then used this configuration as an initial
condition for the full time-dependent dynamical equations
of motion described in the next section. The total kinetic energy
during the time evolution 
was never more than one part in $10^7$ of the soliton energy.
This confirms that the relaxation algorithm has indeed converged
to a static solution to the full equations of motion.}
The soliton configuration we find is a hedgehog configuration,
as at $\lambda\rightarrow\infty$. However, when $\lambda$ is finite,
$\vec \phi \cdot \vec \phi \neq 1$.  The soliton we find 
can be written in the form
\begin{equation}
\vec{\phi}(r,\theta) =\sigma(r) \left( \begin{array}{c}
                                        \sin f(r) \cos \theta\\ 
                                        \sin f(r) \sin \theta\\ 
                                        \cos f(r)
                                \end{array} \right)
\label{solitonform}
\end{equation}
with $f(r)$ satisfying the same boundary conditions as 
above.\footnote{Note 
that we could have rewritten the static Hamiltonian in terms
of $\sigma(r)$ and $f(r)$, discretized that Hamiltonian in $r$, 
and then used a conjugate
gradient algorithm to find these two functions of $r$. This
would have been less computationally intensive than finding 
$\phi^i(x,y)$ as we did.  However, the expressions we obtain
by varying our static Hamiltonian relative to the fields $\phi^i$ 
at each lattice site,
and indeed the results we obtain for $\phi^i$ at each lattice
site in a soliton configuration, 
are precisely what we need in the next section when
we analyze 
soliton--soliton collisions, which are of
course not circularly symmetric and so cannot be written
in terms of $\sigma(r)$ and $f(r)$.}
We depict the soliton configuration in 
Fig. \ref{fig:solitonprofiles}.

After obtaining a baby skyrmion at $\lambda=15$, we used
the resulting configuration as the initial condition for relaxation
at $\lambda=14$, and so found the soliton configuration 
at this $\lambda$.  We repeated this process step-by-step in $\lambda$,
finding solitons for values of $\lambda$ down to $\lambda=8$.
At $\lambda=7$, energy minimization led to 
a configuration with zero energy, instead of to a soliton. 
We then used the $\lambda=8$ soliton as an initial configuration
for relaxation at $\lambda=7.9$, and so on down to $\lambda=7.6$
where again no soliton was found.  We therefore know that 
a stable soliton exists at $\lambda=7.7$. It is a logical
possibility that there is a stable soliton at $\lambda=7.6$
even though our relaxation algorithm did not find one. We think this is
unlikely, because the soliton configurations which we have
found at $\lambda=7.7$ and $\lambda=7.8$ are very
similar, and we therefore believe that the $\lambda=7.7$ soliton
is a very good starting configuration from which to
find the $\lambda=7.6$ soliton if it existed. We therefore
conclude that classically stable solitons exist only for 
$\lambda>\lambda_c$,
with $7.6 < \lambda_c < 7.7$.

\begin{table}
\begin{center}
\begin{tabular}{|r|r|} \hline 
$\lambda$ & Energy \\ 
\hline\hline 
15 & 19.1792\\ 
\hline 
14 & 19.1503 \\ 
\hline 
13 & 19.1161 \\ 
\hline 
12 & 19.0751 \\ 
\hline 
11 & 19.0250 \\ 
\hline 
10 & 18.9618 \\ 
\hline 
9 &  18.8791 \\ 
\hline 
8 & 18.7619 \\ 
\hline 
7.9 & 18.7473 \\
\hline
7.8 & 18.7309 \\ 
\hline
7.7 & 18.7131 \\ 
\hline
7.6 & no soliton \\
\hline
\end{tabular}
\caption{Energy of Static Solitons at various Lambdas.}
\label{tab:lamener}
\end{center}
\end{table}
In Table I, we give the energies of the solitons which we have
found for various values of $\lambda$. In Fig. \ref{fig:solitonprofiles}, 
we depict the field configuration and energy density for the solitons
we have obtained for $\lambda=15$ and $\lambda=7.7$.  We note
that even though $\lambda=7.7$ is only just above $\lambda_c$,
the soliton configuration does not look very different from
that at much larger values of $\lambda$, and the 
soliton energy is also little changed.  Note that 
the deviation from $\sigma(r)=1$ 
is only at most $20\%$ for a soliton with $\lambda=7.7$ which
is on the edge of instability.
The central energy density does increase
by almost a factor of two as $\lambda$ is reduced from 15 to 7.7.
Note, however, that the total energy is almost unchanged, and
actually decreases slightly.  The soliton radius decreases
as $\lambda$ is reduced towards $\lambda_c$, but does not
decrease dramatically. 
The definition of $R_{\rm sol}$ is of course somewhat arbitrary;
if we take it to be the radius
inside which $90\%$ of the total energy of the soliton is found,
we find $R_{\rm sol}=3.31$ for $\lambda=15$ and $R_{\rm sol}=2.83$
for $\lambda=7.7$.


Although the energy density
and $\sqrt{ \vec\phi\cdot\vec\phi}=\sigma$ are circularly
symmetric,
the fields $\phi^1$ and $\phi^2$ in a 
soliton configuration are not circularly symmetric.
If we only observed a single static soliton, this would
be of no consequence: in a hedgehog configuration, the different
possible choices for $\phi^1$ and $\phi^2$ are related
simply by rotations in space. However, 
when we describe a configuration of two well-separated solitons in
the next Section, the relative 
angle $\alpha$ between their orientations
does matter. That is, specifying such a configuration
requires giving the relative position and velocity of the
centers of the two solitons and the angle $\alpha$.
The first soliton 
in such a configuration can be mapped onto the second by a translation
followed by a rotation by an angle $\alpha$ about the soliton center.

\section{Colliding Solitons}

With solitons in hand, we are ready to study what happens when
they collide.  For this purpose, we need discretized equations
of motion and a numerical algorithm to evolve an initial
configuration, now specified by $\phi^i$ and $\dot\phi^i$ at 
each lattice site, forward in time.  
We begin by writing a discretized Lagrangian which is a function
of $\phi^i$ and $\dot\phi^i$ at each of the lattice sites,
at a single time $t$.
We discretize the time-independent terms as described in
the previous Section.  There are no spatial derivatives of 
$\dot\phi^i$ in the Lagrangian, so discretizing terms
involving $\dot\phi^i$ is trivial.  We then use the Euler-Lagrange
procedure on this Lagrangian written in terms of $3\times 125 \times 125$
$\phi$'s and $3\times 125 \times 125$ $\dot\phi$'s, and obtain equations
of motion which specify the
$3\times 125 \times 125$ $\ddot\phi$'s.
These equations of motion take the form of
three coupled linear equations for $\ddot \phi^1$, $\ddot \phi^2$ and 
$\ddot \phi^3$ at a given lattice site, which are easily solved.
We now have an expression for $\ddot \phi^i(t,x,y)$
written in terms of the values of $\phi^i$ and $\dot\phi^i$
at lattice sites within two spatial links
of the site of interest, all at the same time 
$t$.\footnote{Note that because of the way we discretize 
spatial derivatives in the Lagrangian,
expressions in the equations of motion with
mixed time-space derivatives 
such as $\partial_t\partial_x\phi^i$ end up
discretized as $[\dot\phi^i(x+\Delta x,y)-
\dot\phi^i(x-\Delta x,y)]/2\Delta x$.}  
We are now ready to take a step forward in time.

We evolve the system forward in time using the Runge-Kutta-Feldberg
algorithm and the adaptive algorithm of Ref. \cite{Press} for choosing
the size of the time step $\Delta t$.  That is, we first use 
the fifth-order Runge-Kutta-Feldberg algorithm 
to obtain $\phi^i$ and
$\dot\phi^i$ at time $t+\Delta t$.  This fifth-order method is special
because a rearrangement of the fifth-order function evaluation terms
results in a fourth-order Runge-Kutta expression.\footnote{This hidden
fourth-order expression is referred to as an {\it embedded}
Runge-Kutta formula due to the fact that it can be obtained with no
additional function evaluations.}  We then have two different
estimates (fourth order and fifth order)
for $\phi^i$ at $t+\Delta t$ at each lattice site, and can
evaluate the discrepancy between the two estimates for each of the
$3\times 125\times 125$ $\phi$'s and $\dot\phi$'s.  
If the largest discrepancy is
larger than a specified tolerance, we reject the step and begin anew
with a smaller $\Delta t$. We use the largest discrepancy to estimate
how much $\Delta t$ should be reduced.  If all discrepancies are
smaller than the specified tolerance, we accept the result of the
fifth-order calculation for $\phi^i$ and $\dot\phi^i$
at time $t+\Delta t$. 
After a successful step forward in time, 
we use the largest discrepancy (which must have been less than the tolerance
since the step forward was accepted)
to estimate by how much we can
safely increase $\Delta t$ when we take our next step forward in time.
In the simulations of collisions which we describe below,
the tolerance is 
such that the timestep selected by the
adaptive algorithm is approximately
$0.01 \lesssim \Delta t \lesssim 0.05$.  
Note that we do {\it not} use conservation of energy as our
criterion for acceptance or rejection of a step forward in
time.  This makes it fair to use a check of the conservation of
energy as an independent measure of the accuracy of our evolution
algorithm. We do this at various points below.

We choose fixed boundary conditions, with $\vec\phi$ fixed to its
vacuum value $(0,0,\sigma_{\rm vac})$ at the boundaries of our $125\times 125$
grid, where $\sigma_{\rm vac}$ solves 
$(\sigma_{\rm vac}^2-1)\sigma_{\rm vac}=\frac{\mu^2}{4\lambda}$ 
and is $\sigma_{\rm vac}\simeq 1+\frac{\mu^2}{8\lambda}$ for large $\lambda$.
Since the solitons have radii of order $R_{\rm sol}\simeq 3$, we choose
initial conditions with two solitons whose centers are a distance $10$
apart.
We initialize $\vec \phi$ by adding these two soliton configurations.
(That is, we take $\vec\phi_{\rm vacuum} + (\vec\phi_{\rm
first~soliton}-\vec\phi_{\rm vacuum}) + (\vec\phi_{\rm
second~soliton}-\vec\phi_{\rm vacuum})$.)  The resulting configuration
is not precisely a minimum of the static Hamiltonian, but the two
solitons are far enough apart that this is not a big concern. To
obtain a soliton moving with an initial speed $v$ in the positive $x$
direction, we simply initialize 
\begin{equation}\dot\phi^i(x,y)=-v[\phi^i(x,y)-
\phi^i(x-\Delta x,y)]/\Delta x
\label{boost}
\end{equation}
at time zero.  For simplicity, we are
using a Galilean boost. This is appropriate for $v\ll 1$. When we use
this prescription with a velocity at which relativistic corrections
are becoming important, the initial condition we have specified is not
the correct Lorentz-boosted, Lorentz-contracted soliton. In this
circumstance, as the system is evolved forward in time, the soliton
radiates some energy and quickly settles down to become a (correct)
relativistic soliton moving with a velocity somewhat less than $v$.
For example, when we set $v=0.8$ in our Galilean boost prescription
for the initial condition, we in fact end up with a soliton moving at
a speed of $0.61$.

\begin{figure}[t]
\begin{center}
\epsfig{file=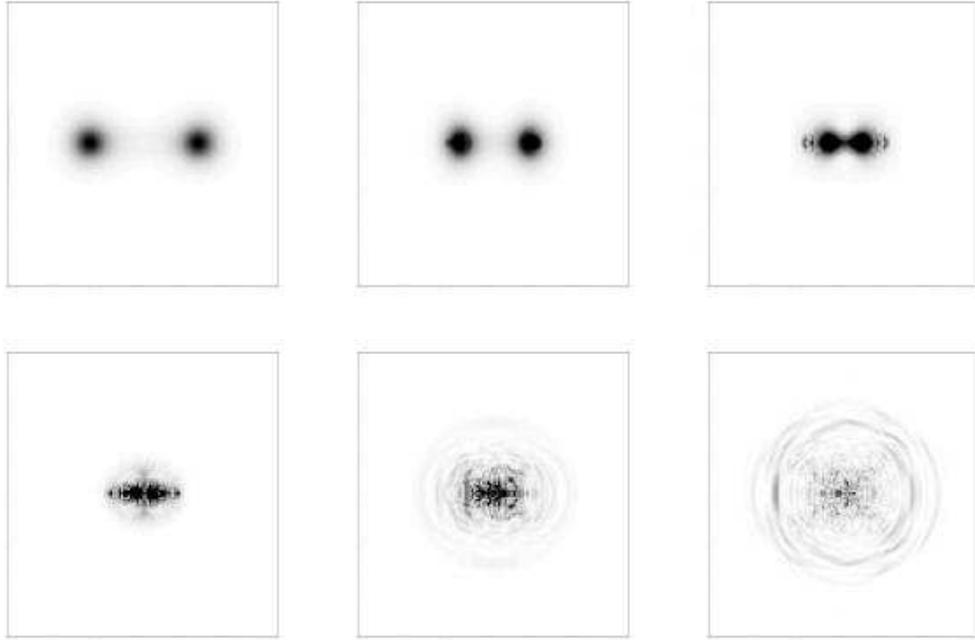,width=5.5in}
\end{center}
\caption{Sequence of snapshots of the energy density during a
collision between two solitons which results in the destruction of
both. The grey scale indicates energy density.  In this simulation,
$\lambda=10$, the initial velocity of each soliton is $v=0.5$, the
impact parameter is $b=0$, and the solitons have a relative
orientation angle $\alpha=0$ in the initial configuration.  
The images are at times $t=0,4,8,12,16,20$.
In this and in all subsequent figures showing soliton--soliton collisions,
each panel shows a $25\times 25$ square (in our units in
which $\kappa=1$) and the initial separation between solitons
is 10. The lattice spacing is $\Delta x=0.2$.} 
\label{fig:firstcollision}
\end{figure}
\begin{figure}[t]
\begin{center}
\begin{tabular}{cc}
\epsfig{file=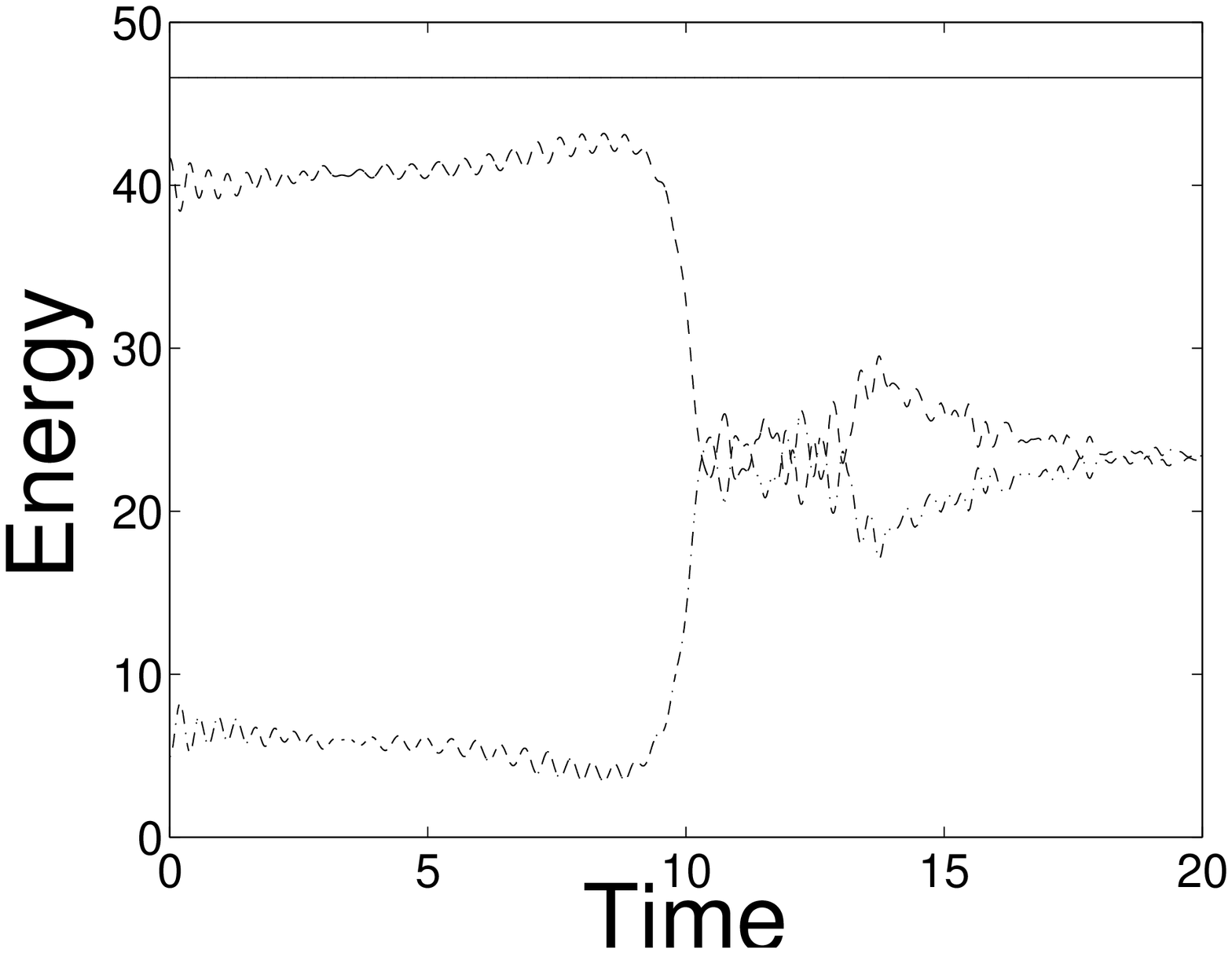,width=2.7in} & 
\epsfig{file=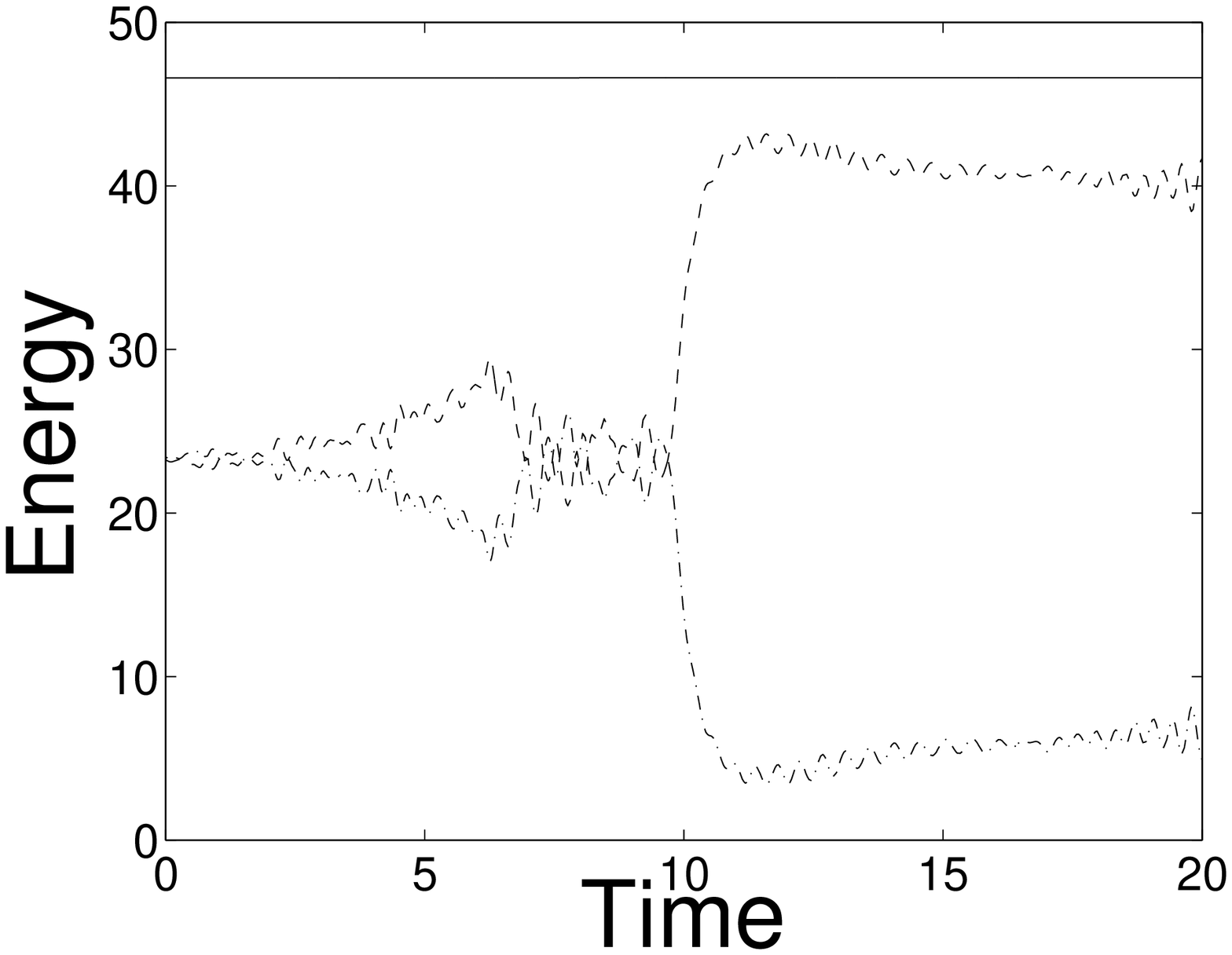,width=2.7in}
\end{tabular}
\end{center}
\caption{Left panel: Kinetic, potential and total energies during
the soliton--soliton collision shown in Fig. \ref{fig:firstcollision}
with $\lambda=10$ and $v=0.5$.
The topmost curve (constant to better than two parts in $10^5$) is
the total energy.  Of the other two curves, the one that begins
low is the kinetic energy, the one that begins high is the
potential energy, including spatial gradient energy.
Right panel: Same, during the time--reversed evolution. We
reverse the sign of all $\dot\phi^i$ in the final configuration
of Fig. \ref{fig:firstcollision}, 
and then watch the evolution algorithm recreate the 
initial configuration of Fig. \ref{fig:firstcollision}.
}
\label{fig:threeenergies}
\end{figure}
We begin by analyzing collisions between two solitons 
in the theory with
$\lambda=10$. We choose initial conditions in which both
solitons are moving (towards each other) with velocity
$v=0.25$, with zero impact parameter. We choose an initial relative 
orientation angle $\alpha=0$,
meaning that
one soliton is obtained from the other by translation without
rotation.  Previous work shows that two static solitons 
with this relative orientation repel each other.\cite{Piette3}
This is consistent with what we find: for low velocities, as
for example for $v=0.25$, the two solitons bounce off each other
and return whence they came.  We now increase $v$ to $0.5$.
This time, the outcome, depicted in Fig. \ref{fig:firstcollision},
is that the solitons are destroyed in the collision. The final
state is a cloud of debris, namely small amplitude oscillations
of the $\vec\phi$ field spreading outwards from the scene
of the collision.  
In Fig. \ref{fig:threeenergies}, we show the kinetic energy, potential
energy and total energy for the collision shown in 
Fig. \ref{fig:firstcollision}. (By ``potential energy'' we mean the 
contribution
to the energy from all those terms in the Hamiltonian with no time
derivatives. Most of this energy is due to spatial gradients of
the fields.)  First, we see that the total energy is conserved,
in fact to better than two parts in $10^5$.  The kinetic energy is
not zero initially, because the solitons are moving. As the solitons
approach each other, the kinetic energy decreases. This confirms
that the interaction is repulsive: the solitons
slow down and deform as they approach. As the solitons approach
each other more closely, at some point their
deformation becomes sufficient that they are no
longer stable, and they fall apart.
The resulting outgoing waves have approximately equal kinetic
and potential energy, as expected for traveling waves.
It is quite clear from Fig. \ref{fig:threeenergies}, if it was not already
clear from Fig. \ref{fig:firstcollision}, that the solitons
have been destroyed.

As a stringent check of the accuracy of
our time evolution algorithm, we take the final configuration
from our simulation, reverse the sign of $\dot\phi^i$, and
evolve it for the same period of time as we did initially. The
second panel of Fig. \ref{fig:threeenergies} shows the behavior
of the energies during this 
``backwards-in-time'' evolution.  It is clear that the debris
reconstitutes itself into two solitons!  The sequence of 
snapshots of the energy density looks almost exactly like
those in Fig. \ref{fig:firstcollision}, but in the
opposite order in time. The discrepancies between
the energy density in the initial configuration and that in 
the configuration obtained after soliton collision and destruction 
followed by time-reversed evolution and soliton recreation
differ by at most $1/40$ of
the energy density at the center of the soliton. The total energy
is conserved to better than one part in $10^4$.  

As a further check of the stability of our algorithm, we have also simulated
soliton--antisoliton annihilation. We obtain an antisoliton
configuration from a soliton configuration by making
the transformation $\phi^2\rightarrow-\phi^2$, equivalent
to taking $\theta\rightarrow -\theta$ in (\ref{guess}) or
(\ref{solitonform}). This turns a hedgehog configuration
into an anti-hedgehog configuration, and hence yields an antisoliton.
We find that analyzing soliton--antisoliton collisions
using our evolution algorithm is no more difficult than analyzing
soliton--soliton collisions.  
We were able to follow the annihilation process
with energy conserved to better than one part in $10^4$.  Now, with
confidence in the accuracy and stability of our evolution algorithm,
we proceed to analyze the outcome of soliton--soliton collisions with
a variety of initial conditions.  

\begin{figure}[t]
\begin{center}
\epsfig{file=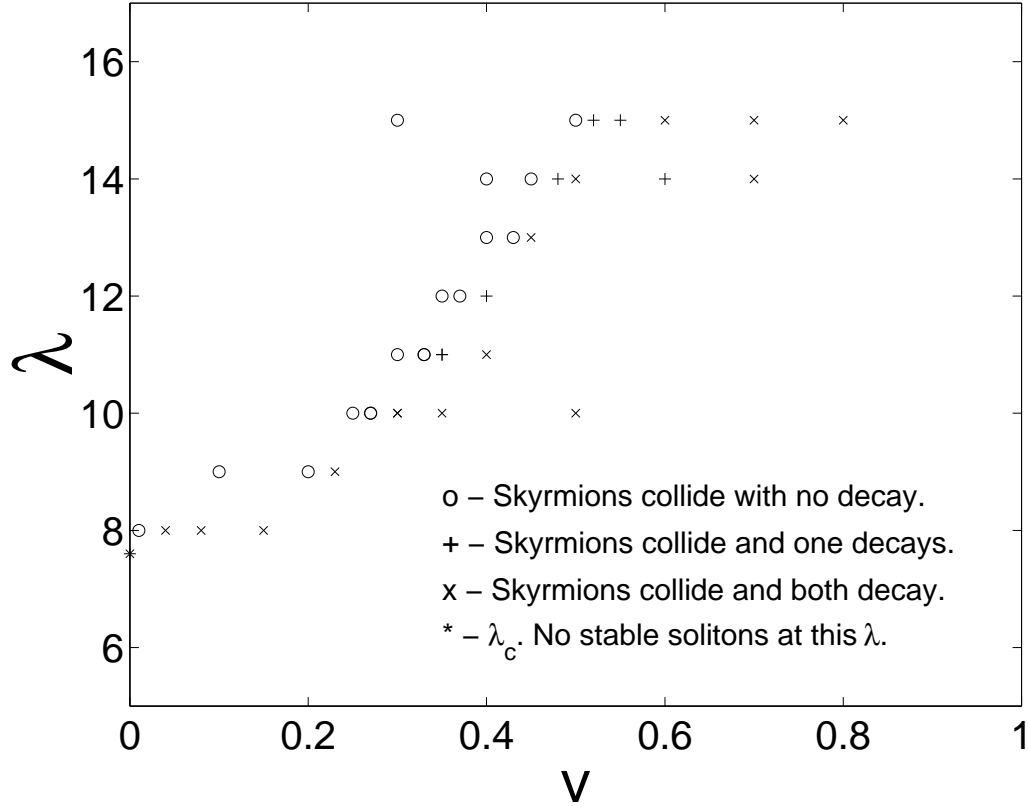,width=5.5in}
\end{center}
\caption{Outcome of soliton--soliton collisions with different
initial velocities and different values of the parameter $\lambda$.
All collisions have impact parameter $b=0$ and relative orientation
angle $\alpha=0$.
Note that $v$ is the velocity parameter in (\ref{boost}).
If $v$ is large enough that relativistic effects are significant,
the actual velocity of the soliton is somewhat less than $v$.
For example, $v=0.8$ yields a soliton with velocity $0.61$.}
\label{fig:lamvel}
\end{figure}
We first explore how the outcome of a collision depends on 
$\lambda$ and $v$, keeping the impact parameter $b=0$ and
the relative orientation angle $\alpha=0$ as above. The results of many
simulations are summarized in 
Fig.~\ref{fig:lamvel}.
We discover that for any $\lambda$, there is a critical velocity
$v_c$ below which the solitons rebound without decaying, and
above which one or both (usually both) solitons are destroyed.
This critical velocity goes to zero as $\lambda\rightarrow \lambda_c$.
As $\lambda$ is increased, $v_c$ increases, reaching about half
the speed of light 
for $\lambda$ about twice $\lambda_c$

\begin{figure}[t]
\begin{center}
\epsfig{file=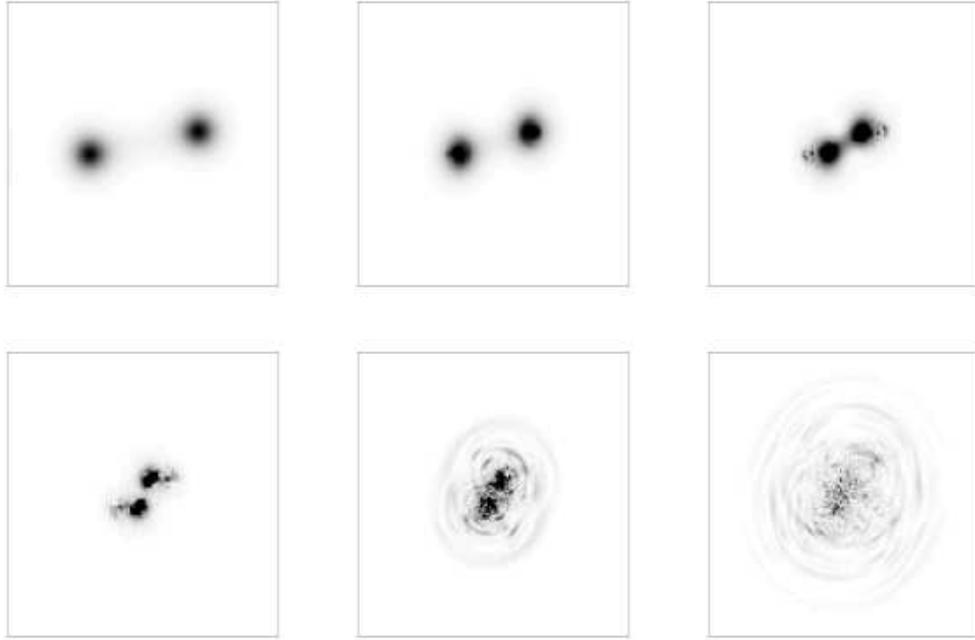,width=5.5in}
\end{center}
\caption{Snapshots of energy density during a collision between two
solitons with impact parameter $b=2.0$ in the theory with $\lambda=10$.
The relative orientation angle is $\alpha=0$.
The initial velocity $v=0.5$ is large enough that the solitons
are destroyed. The time between images is $4.0$.}
\label{fig:nonzerob}
\end{figure}
We now return to $\lambda=10$, $v=0.5$, still keeping $\alpha=0$ 
and ask how the outcome
of a collision depends on the impact parameter $b$. For
$b=2.0$, both solitons decayed into traveling waves, as
we found for $b=0$ above. We show the outcome of this collision
in Fig. \ref{fig:nonzerob}.
Note that $b=2.0$ is a 
substantial impact parameter, comparable to the 
soliton radius $R_{sol}\simeq3$.
We find that the solitons still decay if $b=3.2$.
An impact parameter $b=4.0$, however, yields
a collision which is sufficiently peripheral that the solitons emerge intact,
deflected from their initial directions of motion
by about $45^\circ$. We can describe our results by saying
that the critical velocity $v_c$ above which soliton decay
is the outcome of the collision increases with increasing impact parameter.
For $b=0$, Fig. \ref{fig:lamvel} shows that $0.27 < v_c < 0.3$.
We now see that $v_c=0.5$ for a nonzero impact parameter in
the range $3.2 < b < 4.0$.  We have also done several more simulations
with $b=2.0$ and various initial velocities, and find that
for $b=2.0$, the critical velocity is $0.3<v_c<0.4$.
We conclude that soliton decay
does not require collisions with small or finely-tuned impact parameters.
Although increasing $b$ from zero increases the critical velocity
$v_c$ required to destroy the solitons somewhat, 
it remains easy to destroy solitons as long as 
the impact parameter is less than or comparable to
the soliton radius.

\begin{figure}[t]
\begin{center}
\vspace{-0.1in}
\epsfig{file=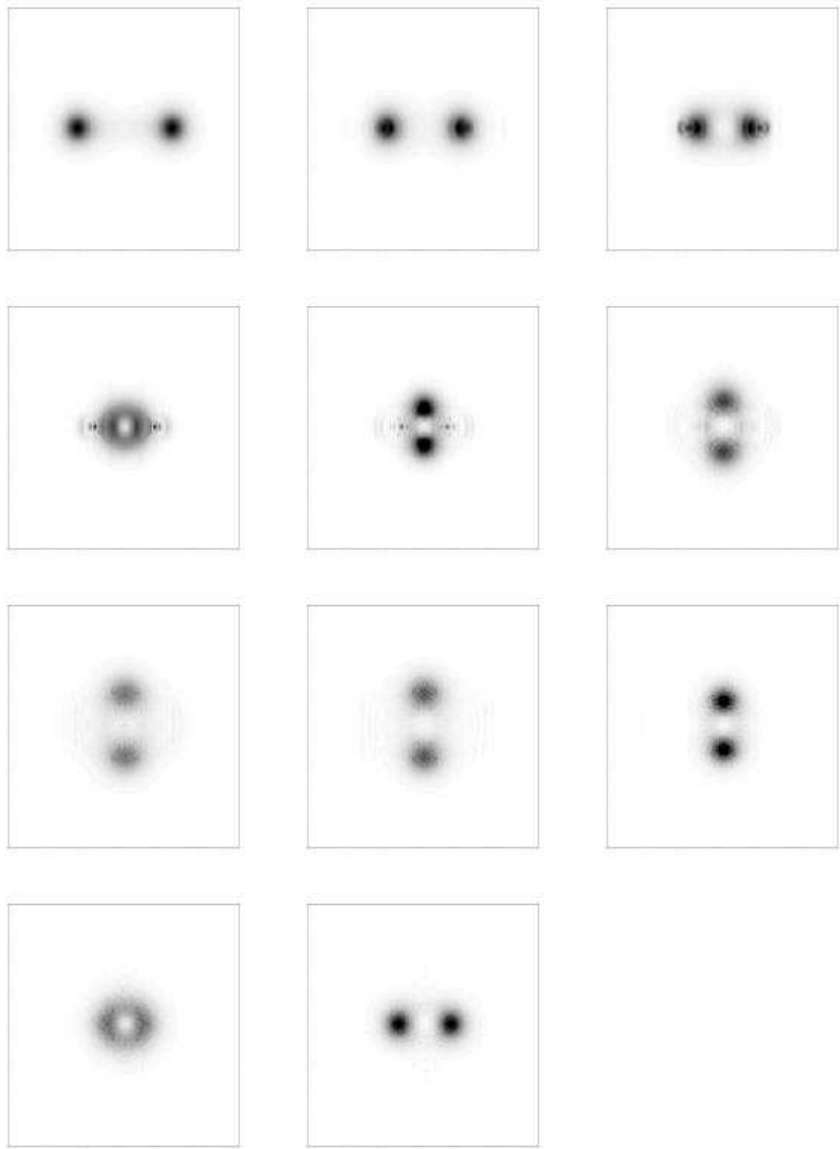,width=4.7in}
\vspace{-0.3in}
\end{center}
\caption{Snapshots of energy density during a collision between 
solitons with relative orientation $\alpha=180^\circ$,
impact parameter $b=0$, and initial velocity $v=0.25$ in the theory
with $\lambda=10$. The solitons are not destroyed and 
(eventually) form a
classically stable bound state.  The time interval between images
varies: the images are
at times
$t=0,4,8,12,16,20,24,28,34,42,50$.} 
\label{fig:orientationpi}
\end{figure}
All the collisions we have described to this point have had
the same relative orientation. For $\alpha=0$, 
low velocity collisions yield a rebound, in 
which each soliton reverses direction, while
higher velocity collisions lead to soliton destruction.
We now consider a collision (with $\lambda=10$, $v=0.25$ and $b=0$)
between two solitons with a relative orientation angle 
$\alpha=180^\circ$.
That is, the second soliton in the initial configuration is
obtainable from the first by a translation and a $180^\circ$
rotation. The interaction between static solitons with
this orientation is known to be attractive.\cite{Piette3}
We show the outcome of a low velocity collision in
Fig. \ref{fig:orientationpi}.  The work of Ref. \cite{Piette2}
reveals that in the $\lambda\rightarrow\infty$ theory, there
is a stable, ring-shaped, soliton with winding number 2.  It appears
that the final state of the collision in  Fig. \ref{fig:orientationpi}
will be a soliton of this form, although it will differ in its details
from that of Ref. \cite{Piette2} 
since $\lambda$ is finite. What we observe in Fig. \ref{fig:orientationpi}
is that the incident solitons at first
scatter by $90^\circ$, but then do not escape to infinity.
They fall back upon one another, and rescatter by $90^\circ$.
There are small outgoing ripples at late time, but they have
too little energy density to be visible in Fig. \ref{fig:orientationpi}.
We expect that were we to run the simulation for a long time,
in a big enough box that outgoing ripples never return, 
we would see repeated $90^\circ$ scatterings, with
the solitons escaping less and less far away each time, all the
while radiating small outgoing ripples, and eventually settling
down to become the static, ring-shaped configuration.

\begin{figure}[t]
\begin{center}
\epsfig{file=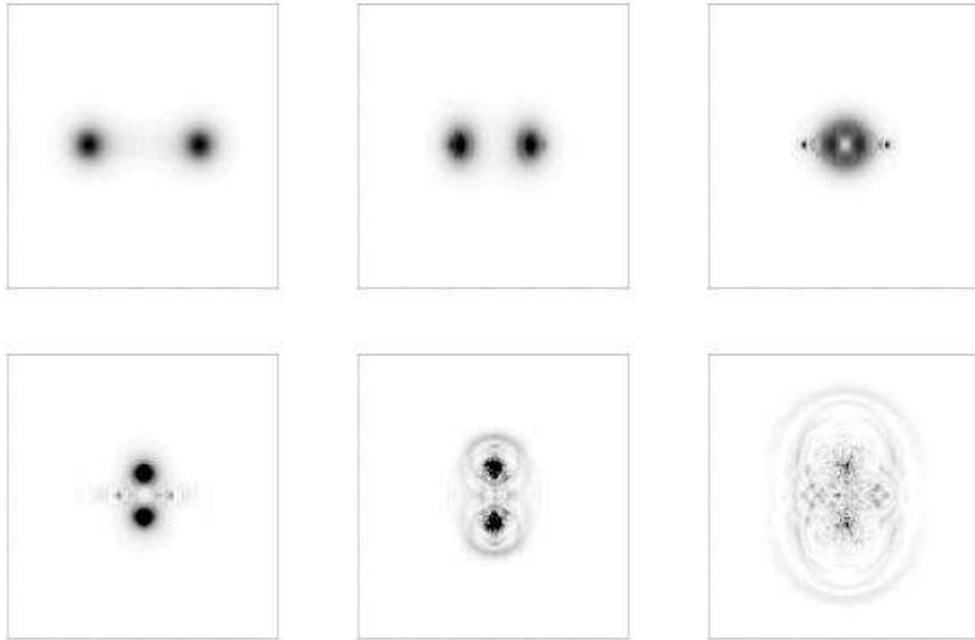,width=5.5in}
\end{center}
\caption{Snapshots of energy density during a collision between
two solitons 
with relative orientation $\alpha=180^\circ$
in the theory with $\lambda=10$. The impact
parameter is $b=0$.  The initial velocity is large enough ($v=0.5$)
that the two solitons decay.  The time between each image is $4.0$.}
\label{fig:decayorientpi}
\end{figure}
As we increase the incident velocity, we find that for $v>v_c$ 
with $0.43<v_c<0.48$, the outcome of the collision is soliton
destruction rather than $90^\circ$ degree
scattering followed by the formation of a bound state.
We show an example of collision induced decay in a collision
with relative orientation $\alpha=180^\circ$ in Fig. \ref{fig:decayorientpi}.
Note that the critical velocity above which soliton
destruction is the outcome is somewhat larger than, but still
comparable to, that we found previously
for $\alpha=0$. We have not mapped out 
$v_c$ vs. $\lambda$ for the $\alpha=180^\circ$ orientation as we did
in Fig. \ref{fig:lamvel}, but we expect that the
figure would be qualitatively similar.
One new feature, though, would be that at large $\lambda$ there
would be two different outcomes possible for 
collisions with $v<v_c$: bound state formation
(for low enough $v$) and $90^\circ$ scattering followed by
the escape of the two intact solitons to infinity (for larger $v$ 
which is still less than $v_c$).  At $\lambda=10$, 
we do not find any velocities
for which  $90^\circ$ scattering followed by escape occurs. 
It must occur at larger $\lambda$, since
it certainly occurs at large enough velocities for $\lambda\rightarrow\infty$,
when $v_c\rightarrow 1$. 

The collision shown in Fig. \ref{fig:decayorientpi}
is an example of a simulation in which
the initial velocity ($v=0.5$ in this case) is only just above
the critical velocity ($0.43<v_c<0.48$ in this case). 
In this circumstance, what we generically
observe is that the solitons scatter, separate a little, but are
sufficiently distorted as a result of the scattering
that after separating a little they
fall apart.  We observe this phenomenon also at $\alpha=0$,
except in this case the solitons scatter by bouncing 
back in the direction whence they came, then separate
a little, and then fall apart.  
At velocities which 
are somewhat larger than $v_c$, as for example in the
collision shown in Fig. \ref{fig:firstcollision}, we find that
soliton destruction occurs more promptly, during the
initial collision.

We now consider collisions between solitons with a relative
orientation angle $\alpha=90^\circ$, still 
with $\lambda=10$ and $b=0$.  For this 
relative orientation, there is no force between
static solitons.\cite{Piette3}  We find the same possible
outcomes as we did for $\alpha=180^\circ$.
As a function of increasing velocity, the outcome of a collision
is either capture to form the ring-shaped bound state, 
or soliton destruction. (Again, 
scattering 
by an angle of $90^\circ$
followed by the escape of two
intact solitons would be a possibility at larger $\lambda$.)  
The critical velocity above which soliton decay occurs
is $0.25<v_c<0.3$.  

\section{Concluding Remarks}

We have analyzed soliton--soliton collisions in a $2+1$-dimensional
theory with metastable baby skyrmion solutions.  We find 
classically stable soliton
solutions for values of the parameter
$\lambda$ which are larger than
$\lambda_c\sim 7.6$.  These solitons are prevented from decaying
by a finite energy barrier and so can decay if supplied
with sufficient energy, for example in a collision with
a second soliton.  We have mapped out the space of initial
conditions under which the outcome of a soliton--soliton
collision is the destruction of one or both solitons.  
We find that soliton decay results whenever two solitons
collide with
an incident velocity greater than
some $v_c$.  This critical velocity depends on the
parameters in the problem.  It goes to zero as $\lambda\rightarrow 0$
and the solitons cease to be classically stable. It goes 
to the speed of light as $\lambda\rightarrow\infty$ and the barrier
to decay becomes infinite.  However, $v_c$ does not rise
particularly rapidly with $\lambda$: with other
parameters chosen as in Fig. \ref{fig:lamvel}, $v_c$ is only half the
speed of light for $\lambda\sim 2 \lambda_c$.  
Thus, soliton destruction does {\it not} require that the theory
have a value of $\lambda$ lying in some narrow range
just above $\lambda_c$.
The impact parameter $b$ need not
be finely tuned either.  Not surprisingly, $v_c$ is lowest
for collisions with $b=0$.  However, $v_c$ increases
by less than a factor of two for $b$ of order the soliton
radius.  $v_c$ also depends on the relative orientation 
angle $\alpha$ between the two solitons in the initial state.
Here too, the dependence is weak.  In the example we explored
in detail, we found that as $\alpha$ changes from $0^\circ$
to $180^\circ$, $v_c$ varies between $0.25<v_c<0.3$ and $0.43<v_c<0.48$.
Thus, although $v_c$ does depend on $\lambda$ and on the 
parameters other than the velocity needed to fully specify
a choice of initial
conditions, the variation of $v_c$ is not dramatic.  
Soliton decay is not restricted to specially chosen velocities,
impact parameters, orientations, or values of $\lambda$.
Soliton decay is a generic outcome of soliton--soliton collisions.

Our findings motivate future investigation of collisions between
metastable solitons in the $3+1$-dimensional electroweak theory.
Previous work on two-particle collisions involving these electroweak
solitons has focussed on collisions between a $W$ boson and a soliton
\cite{FGLR}.  In such collisions, the probability for soliton decay
falls exponentially as the (rough) analogue of $\lambda$ is increased
above the (rough) analogue of $\lambda_c$.  This was traced to two
facts: First, causing one of these solitons to decay requires
delivering sufficient energy to one particular mode of oscillation of
the soliton. Second, a generic incident $W$-boson couples very weakly
to the mode which must be energized if decay is to be induced.  We
find no analogue of this difficulty in our analysis of
soliton--soliton collisions in $2+1$ dimensions.  If there is a
particular mode which must be excited, then soliton--soliton
collisions seem to generically deliver energy to this
mode.
And, we certainly see no evidence of soliton decay being
restricted to theories with $|\lambda-\lambda_c|\ll\lambda_c$.  This
suggests that collisions between two TeV scale particles which can be
modeled as electroweak solitons (rather than between one $W$-boson
and one such particle) may be an arena in which two-particle
collisions generically lead to baryon number violation.  As we
stressed in the Introduction, however, the metastable baby skyrmions
we analyze differ in several important qualitative respects from
metastable electroweak solitons. Furthermore, our analysis has been
purely classical whereas the analysis of $W$-soliton collisions in
Ref. \cite{FGLR} is quantum mechanical.  Although our results motivate
an analysis of collisions between electroweak solitons, they should
not be taken to provide even qualitative guidance as to the outcome of
such a study.

\vspace{3ex}
{\samepage 
\begin{center} Acknowledgements \end{center}
\nopagebreak
We thank S. Bashinsky,
M. A. Halasz, J. Leffingwell, A. Lue, B. Scarlet and S. Sondhi for
very helpful discussions.  
KR is grateful to the Department of Energy's
Institute for Nuclear Theory at the University
of Washington for generous hospitality and support during
the completion of this work.  
The work of KR is also supported in part by the
Department of Energy under cooperative research agreement
DF-FC02-94ER40818, by a DOE OJI Award and by the Alfred P. Sloan
Foundation.}

\end{document}